\documentclass[]{spie}  

\usepackage{longtable}
\usepackage{amsmath,amsfonts,amssymb}
\usepackage{graphicx}

\usepackage[colorlinks=true, allcolors=blue]{hyperref}

\title{The Roman Coronagraph Community Participation Program: corgisim — a simulation suite for the Nancy Grace Roman Space Telescope Coronagraph Instrument}

\author[a]{Jingwen Zhang}

\author[b]{Sophie Noiret}
\author[a]{Maxwell A. Millar Blanchaer}
\author[c]{Jason Wang}
\author[b]{Alexis Lau}
\author[d]{Neil Zimmerman}
\author[e]{Jessica Gersh-Range}
\author[a]{Eric Shen}
\author[f]{Kevin Ludwick}
\author[g]{Taichi Uyama}
\author[h]{Chen Xie}
\author[i]{Vanessa P. Bailey}
\author[i]{John Krist}
\author[b]{Elodie Choquet}
\author[h]{Julien Girard}
\author[h]{Alexis Bidot}
\author[b]{Arthur Vigan}

\affil[a]{Department of Physics, University of California, Santa Barbara, CA 93106, USA}
\affil[b]{Aix Marseille Universit\'{e}, CNRS, CNES, LAM, Marseille, France}
\affil[c]{Center for Interdisciplinary Exploration and Research in Astrophysics (CIERA), Northwestern University, 1800 Sherman Ave, Evanston, IL, 60201, USA}
\affil[d]{NASA Goddard Space Flight Center, 8800 Greenbelt Rd, Greenbelt, MD, USA, 20771}
\affil[e]{DM Telescopes LLC, Raleigh, NC 27615, USA}
\affil[f]{University of Alabama at Huntsville, Huntsville, AL 35805, USA}
\affil[g]{National Astronomical Observatory of Japan, 2-21-1 Osawa, Mitaka, Tokyo 181-8588, Japan}
\affil[h]{Space Telescope Science Institute, 3700 San Martin Drive, Baltimore, MD 21218, USA}
\affil[i]{NASA Jet Propulsion Laboratory, California Institute of Technology, Pasadena, CA 91109, USA}
\authorinfo{Further author information: (Send correspondence to J.Z.)\\J.Z.: E-mail: jwzhang@ucsb.edu}

\begin{document} 
\maketitle

\begin{abstract}
NASA’s Roman Space Telescope will feature a pathfinder Coronagraph Instrument to demonstrate advanced high-contrast imaging from space, paving the way for future missions like the Habitable Worlds Observatory. The Coronagraph Instrument could obtain imaging, polarimetry and spectroscopy of Jupiter analogs in reflected visible light for the first time. We present the development of an open-source simulation package “\texttt{corgisim}” as part of the Roman Coronagraph Community Participate Program. Built on established optical propagation libraries including \texttt{PROPER} and \texttt{CGISim}, \texttt{corgisim} provides a user-friendly, publicly available Python framework for end-to-end simulations of the  Coronagraph Instrument observations. The package produces high-fidelity, format-compliant data for pre-launch calibration, pipeline testing, and community applications such as target selection and observation planning. We will give an overview of \texttt{corgisim}’s infrastructure, functionalities, and current implementation across planned imaging, polarimetry, and spectroscopy modes, including the ability to simulate host stars, injected companions, and extended disks. We will also highlight suitable applications of \texttt{corgisim} and provide guidance on how users can access and employ the software.

\end{abstract}


\section{INTRODUCTION}
\label{sec:intro}

\label{subsec:motivation}
NASA's Nancy Grace Roman Space Telescope will host the Coronagraph Instrument, a pathfinder for high-contrast imaging at visible wavelengths from space \cite{Kasdin2020}. The Coronagraph Instrument is designed to achieve high contrasts ($10^{-8}$) at close angular separations (3 to 20 $\lambda/D$), potentially enabling the  first reflected light images and spectra of Jupiter analogs \cite{Bailey2023}. The technologies demonstrated by the Coronagraph Instrument will provide critical experience and validate key techniques for future flagship missions, such as the Habitable Worlds Observatory, which aims to directly image and characterize Earth-like planets in the habitable zones of Sun-like stars.

The Roman Coronagraph Community Participation Program (CPP) \cite{Savransky2024} has been established to prepare for and execute the technology demonstration observations of the Coronagraph Instrument through several working groups focused on different aspects of the mission \cite{Wolff2024}. Among these, the Data Reduction and Simulations Working Group\cite{Millar-Blanchaer2024} leads the development of the software infrastructure for both data simulation and processing, including two key open-source packages: \texttt{corgisim}\footnote{\url{https://github.com/roman-corgi/corgisim}}, which simulates the Coronagraph Instrument data products, and \texttt{corgidrp}\footnote{\url{https://github.com/roman-corgi/corgidrp}}, which reduces raw observations \cite{Wang2026}.

In this proceeding, we focus on \texttt{corgisim}, which is designed to simulate realistic raw datasets of the Coronagraph Instrument to support pre-launch calibration, observation planning, and the development and validation of the \texttt{corgidrp} data reduction pipeline. To achieve this, \texttt{corgisim} provides a user friendly Python framework that integrates established optical and detector modeling tools, including \texttt{PROPER}\cite{Krist2007}, the Roman preflight optical prescription\footnote{\url{https://sourceforge.net/projects/cgisim/files/}}, \texttt{CGISim}\cite{Krist2023}, and the EMCCD detector simulator \texttt{EMCCD\_Detect}\footnote{\url{https://github.com/roman-corgi/emccd_detect}}. In the following sections, we present an overview of \texttt{corgisim}, including its current architecture, capabilities, and ongoing development. We describe the simulation flow from input parsing and scene construction through optical propagation, detector modeling, and output generation. We then summarize the present implementation for the  Coronagraph Instrument imaging, polarimetric imaging, and spectroscopy modes. Finally, we discuss example applications of \texttt{corgisim}, including pipeline testing, calibration studies, target planning, and community science, and provide guidance for accessing and using the package.





\section{corgisim SIMULATION FLOW}
\label{sec:architecture}

\begin{figure} [ht]
   \begin{center}
   \begin{tabular}{c} 
   \includegraphics[height=10cm]{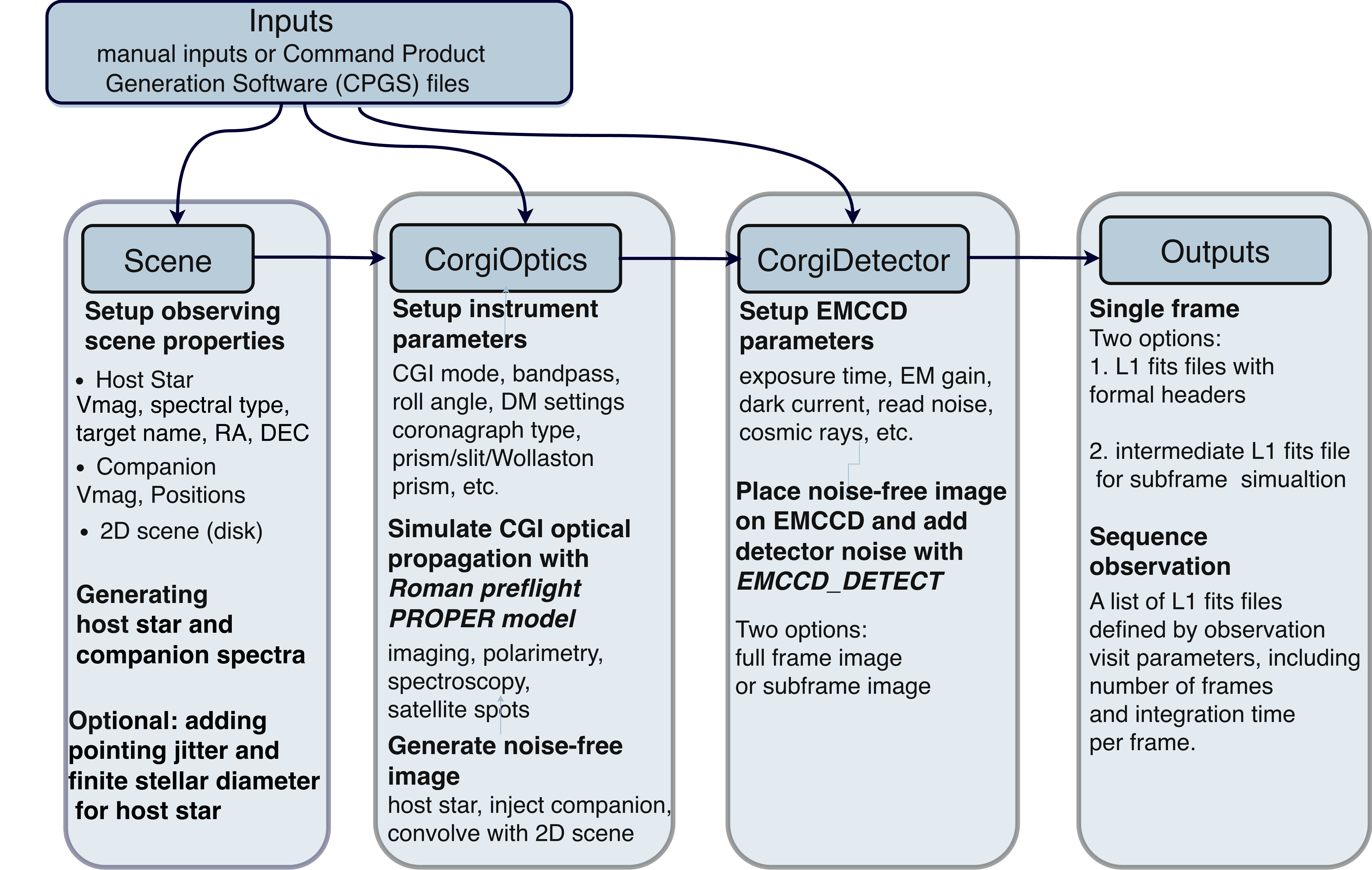}
   \end{tabular}
   \end{center}
   \caption[example] 
   { \label{fig:fig1} 
Flow chart of \texttt{corgisim}}
   \end{figure}

This section introduces the simulation flow implemented in
\texttt{corgisim}, as summarized in Figure~\ref{fig:fig1}. Starting from the inputs, the package defines the
astrophysical scene, propagates the source light through the coronagraph optical
model, places the resulting noiseless image on the EMCCD detector, and writes the
simulated products to FITS files. The following subsections describe the role of
each component in this workflow.

\subsection{Inputs}

\texttt{corgisim} provides two ways for specifying simulation inputs. The first is direct configuration through \texttt{PYTHON}, which gives users flexibility to manually define the astrophysical scene, optical configuration, and detector settings. The second is through Roman's Command Product Generation Software (CPGS) files which contains detailed information for observation setting. CPGS generates XML files which describe an observation scenario. This typically consists of a High Order Wavefront Sensing and Control (HOWFSC) sequence to dig the dark hole (not simulated in corgisim), followed by several visits on a reference star and a target star, with several rolls for each star. When using these XML files as inputs, the user only needs to provide the path to the file. They can also specify a companion with its position and magnitude, which will be added to the scene and simulated along with the target. CPGS inputs only allow the simulation of the Hybrid Lyot Coronagraph (HLC) narrow field of view (NFOV) imaging mode in Band 1 and 4 for now, but future development will include Wide Field of View (WFOV), polarimetry and spectroscopy, as well as the inclusion of satellite spots images.

For both methods, \texttt{corgisim} internally loads the inputs about the astrophysical scene, instrument configuration, and detector parameters through Python dictionaries. The target system is defined by the \texttt{Scene} object, which can include a host star, one or more companions, and an optional two-dimensional (2D) scene such as a circumstellar disk. Host star properties are provided through the \texttt{host$\_$star$\_$properties} dictionary, including the apparent magnitude in V band,  spectral type, target name, RA, DEC, stellar diameter, and a flag indicating whether the source is a science target or a reference star. Companion properties are provided through the \texttt{companion$\_$info} dictionary, with each companion described by its apparent magnitude in V band and relative astrometry with respect to the host star. Users can also provide a custom spectrum and polarization state for each companion; by default, companions are assigned a flat spectrum and are treated as unpolarized.  An extended 2D scene is provided through the \texttt{twoD$\_$scene$\_$info} dictionary. In the current implementation, this dictionary contains the scene contrast relative to the host star, specified as \texttt{contrast} in magnitudes, the path to a FITS image of the disk object, specified as \texttt{disk$\_$model$\_$path}. The FITS image is treated as the intensity map of the extended source, such as a disk, and is assumed to be oriented in the same sky reference coordinate as the host star and sampled at the spatial resolution used by \texttt{corgisim}.  In addition, \texttt{corgisim} can also include finite stellar diameter and pointing jitter effects through the optional \texttt{stellar\_diam\_and\_jitter\_keywords}
input. Table~\ref{tab:scene_inputs} summarizes the input parameters available for the \texttt{Scene} object.

The instrument configuration is handled by the \texttt{CorgiOptics} object. Its primary inputs define the high-level observing setup, including the cgi mode, bandpass, telescope diameter, roll angle, image oversampling, and optional visit metadata. The \texttt{cgi\_mode} parameter determines the simulation mode, such as imaging with \texttt{excam} or spectroscopy with \texttt{spec}, while the bandpass and roll angle define the filter and telescope orientation. More detailed optical settings are provided through the \texttt{optics\_keywords} dictionary, which is passed to the Roman preflight \texttt{PROPER} model \cite{Krist2007,Krist2023} to simulate optical propagation through the  Coronagraph Instrument. These keywords specify the coronagraph type, polarization axis, output image size, use of optical elements such as the focal-plane mask, Lyot stop, field stop, pupil lens, deformable mirrors, and mode-specific components such as prisms and spectroscopy slits. For spectroscopy mode, the \texttt{optics\_keywords} dictionary also defines the slit and prism configuration, as well as the slit position. Additional optional dictionaries can be used to add satellite spots through DM commands or to include finite stellar diameter and pointing-jitter effects. Table~\ref{tab:corgioptics_inputs} summarizes the input parameters available for the \texttt{CorgiOptics} object.

Finally, detector modeling is handled by the \texttt{CorgiDetector} object, which uses the \texttt{emccd\_detect} package to simulate the EMCCD response. Its primary inputs include the \texttt{emccd\_keywords} dictionary and the exposure time for each simulated frame. The \texttt{emccd\_keywords} dictionary specifies detector properties such as EM gain, full-well capacities, dark current, clock-induced charge, read noise, detector bias, cosmic-ray rate, pixel pitch, gain-register size, ADC bit depth, and electrons per data number. These parameters are used to place the noiseless simulated scene onto the detector and generate noisy detector images. Table~\ref{tab:corgidetector_inputs} summarizes the input parameters available for the \texttt{CorgiDetector} object.

\subsection{Optical modeling}

\texttt{corgisim} provides the Coronagraph Instrument imaging, polarimetry, and spectroscopy simulations. The simulation begins by generating the host-star spectrum from the input spectral type, magnitude, and magnitude system. Companion spectra are also generated at this stage, either as default flat spectra normalized to the requested companion brightness or from user-provided spectra. The instrument model then constructs the wavelength-dependent throughput curve for the selected filter, including filter transmission, losses from mirrors, lenses and neutral density along the  Coronagraph Instrument optical path, and the detector quantum efficiency.

For a given cgi mode, bandpass, and coronagraph type, \texttt{corgisim} calls the Roman preflight \texttt{PROPER} model to simulate optical propagation through the coronagraph. The supported configurations include the Hybrid Lyot Coronagraph (HLC) narrow field of view (NFOV) imaging mode, Shaped Pupil Coronagraph (SPC) wide field of view (WFOV) imaging mode, SPC bowtie imaging mode, polarimetric imaging, and SPC spectroscopy \cite{Riggs2021}. In broadband imaging mode, the bandpass is sampled at multiple wavelength points, typically seven for the nominal imaging filters. For the host star, an on-axis coronagraphic point-spread function (PSF) is generated at each sampled wavelength, weighted by the stellar spectrum and system throughput, and integrated over wavelength to produce the broadband residual stellar image in the selected filter. If off-axis companions are provided, they are propagated separately and injected into the simulated scene. Their positions, specified as relative offsets from the host star in sky coordinates, are converted to EXCAM coordinates using the telescope roll angle. The companion images are also simulated at multiple wavelengths across the bandpass, weighted by each companion spectrum and the system throughput, and integrated to produce the broadband companion imaging. If an extended two-dimensional scene is provided, \texttt{corgisim} propagates
it through the instrument model by first normalizing the scene by its total flux.
The normalized scene is then convolved with pre-computed, field-dependent
off-axis point-spread functions sampled on a polar grid. 


In polarimetry mode, \texttt{corgisim} includes the Wollaston-prism configuration \cite{Groff2025} and propagates the corresponding linear polarization states through the optical model. The \texttt{POL0} and \texttt{POL45} settings produce two orthogonal polarization channels, corresponding to 0/90 degree and 45/135 degree analyzer directions, respectively. The polarization signal of the stellar speckle field is computed by configuring the Roman preflight \texttt{PROPER} backend to propagate its built-in Jones pupil model through the coronagraph onto the detector focal plane. Since the Jones pupil is a direct characterization of the system's polarization aberrations, the four output fields at the detector focal plane corresponding to each of the Jones pupil elements can be converted to intensities and selectively combined to obtain the polarized intensity of a specific analyzer channel. This captures the polarization-dependent speckle structure introduced by the  Coronagraph Instrument's optical system. Companion polarimetry is treated separately. Each companion can be assigned an input Stokes vector, \texttt{[I,Q,U,V]}, with respect to the on-sky equatorial coordinate system. The default Stokes vector for a companion corresponds to unpolarized light. \texttt{corgisim} first propagates the companion as an off-axis point source to obtain the total unpolarized intensity at the output. Next, the companion Stokes vector is transformed from the sky frame to the instrument frame via a rotation Mueller matrix corresponding to the telescope roll angle. Finally, the companion Stokes vector in the instrument frame is multiplied by the band-averaged instrument Mueller matrix\footnote{\url{https://roman.ipac.caltech.edu/docs/Roman-Coronagraph-Optical-Model-Mueller-Matrices-450-to-950nm.pdf}} and the Mueller matrices of the selected Wollaston analyzer channels. The companion's total unpolarized intensity is scaled by the resulting normalized intensity in each polarization channel, producing the companion contribution in the 0/90 or 45/135 degree image pair.

In spectroscopy mode, \texttt{corgisim} uses the SPC spectroscopic coronagraph configuration and applies the selected slit and  prism model\cite{Groff2025}. The slit location can be set either in sky coordinates or directly in EXCAM coordinates. If sky-coordinate offsets are provided, \texttt{corgisim} converts them to EXCAM slit offsets using the telescope roll angle. The code then builds a  slit transmission mask from the reference slit-parameter file, applies the requested slit offset, bins the mask to the required sampling, and passes it to the Roman preflight \texttt{PROPER} model as the field-stop array. After the coronagraphic image cube is generated, the optional prism model disperses the image. \texttt{corgisim} loads the prism dispersion solution from a reference file, interpolates the image cube onto a finer wavelength grid, and shifts each wavelength slice accordingly. The dispersed slices are weighted by the source spectrum and system throughput, then summed to form the simulated spectroscopic image. The same framework can be run with no slit or prism, with the slit only, or with both slit and prism, allowing direct comparison of broadband imaging, slit-limited imaging, and dispersed spectroscopic products. Off-axis companions are injected using the same spectral weighting and dispersion treatment, so the final simulated image can include both the stellar speckle field and the companion spectrum. 

In addition, \texttt{corgisim} can optionally add satellite spots to regular
imaging, polarimetric imaging, and spectroscopic simulations by modifying the DM1 voltage map. When \texttt{satspot\_keywords} is provided, the code adds one or two cosine patterns to \texttt{dm1\_v}, with user-defined spot separation in $\lambda/D$, position angle, contrast, and reference wavelength. An optional \texttt{sign} keyword sets whether the cosine pattern is added with positive or negative sign; the default is \texttt{positive}.

\texttt{corgisim} can also include finite stellar diameter and pointing jitter
effects. In this case, the host-star PSF is generated by combining multiple
slightly offset source positions that represent the angular extent of the
stellar disk, or the telescope pointing-jitter distribution. These
weighted images are summed before detector modeling, capturing the PSF
broadening and contrast degradation introduced by finite source size and
pointing instability.

\begin{figure}
\begin{center}
\begin{tabular}{c} 
\includegraphics[height=18cm]{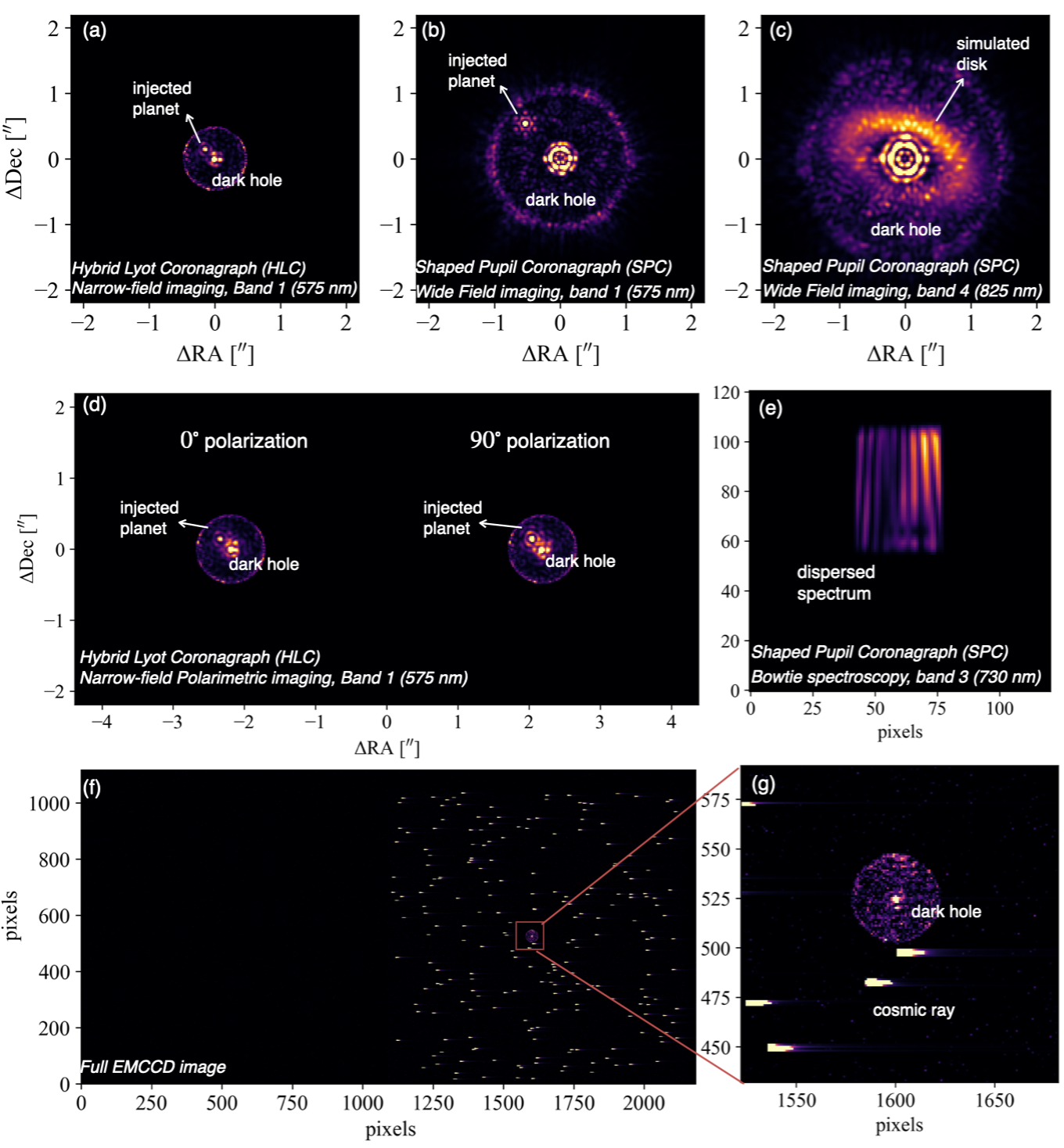}
\end{tabular}
\end{center}
\caption[example] 
{ \label{fig:fig2} Examples of the Coronagraph Instrument simulations generated with \texttt{corgisim}. (a)--(e) Intermediate subframe simulations for different coronagraph designs, bandpasses, and observing modes, including imaging, polarimetry, and spectroscopy of host-star PSFs, injected companions, and circumstellar disks. (f) Full-frame EMCCD simulation incorporating dark current, flat-field variations, cosmic-ray events, and read noise. The output is saved as a standardized FITS file ready for processing by the data-reduction pipeline. (g) Zoom-in view of the dark-hole region.}
\end{figure}


\subsection{Detector simulation}

\texttt{corgisim} uses the \texttt{emccd\_detect} package to simulate detector effects through the \texttt{CorgiDetector} class. The noiseless image from \texttt{CorgiOptics} is first combined from the available scene components. For non-polarimetric simulations, if the simulated image is smaller than the 1024$\times$1024 EXCAM science frame, it is placed onto the detector at the selected location (\texttt{loc\_x} and \texttt{loc\_y}). For polarimetric simulations, the simulated product contains two images corresponding to the two polarization channels. \texttt{CorgiDetector} places these two images onto the EXCAM frame at their expected separated detector locations. For the 0/90 degree configuration, the two images are placed symmetrically along the detector $x$ direction; for the 45/135 degree configuration, they are placed symmetrically along the diagonal direction. After the noiseless scene is placed on the detector frame, \texttt{CorgiDetector} calls \texttt{emccd\_detect} with the exposure time and EMCCD settings to generate the final detector image. This step applies detector effects \cite{Morrissey2023} such as EM gain, dark current, clock-induced charge, read noise, cosmic rays, bias, K-gain, flat field effects, read-out smearing,  and digitization into detector counts.

\subsection{Output Products}

\texttt{corgisim} provides two types of output products. First, intermediate images are stored in the \texttt{SimulatedImage} object as Astropy HDUs, including the noiseless host-star PSF, injected companion image,  2D scene image, and detector image. These products include header comments that record the main simulation settings, and they can be saved as standard FITS files with user-defined filenames.

Second, full-frame detector simulations can be written as Level 1(L1) FITS products. These files contain a primary HDU with global metadata and an image extension with the simulated detector frame. The headers are populated with exposure, detector, observing-mode, roll-angle, polarization, slit/prism, satellite-spot, and instrument-mechanism keywords, following the Coronagraph Instrument header format.  L1-style files are automatically named using the Roman official convention, as \texttt{cgi\_<VISITID>\_<time>\_l1\_.fits}.

\texttt{corgisim} can also generate outputs as an observation sequence consisting of multiple frames rather than a single fits file. In \texttt{observation.py}, the function \texttt{generate\_observation\_sequence()} first propagates the scene through \texttt{CorgiOptics} to create the noiseless host-star PSF and, when present, injected companion components. It then calls \texttt{CorgiDetector} repeatedly to generate the requested number of detector frames for a given exposure time. Each frame is returned as a separate \texttt{SimulatedImage} object with its own detector image and FITS metadata. For CPGS-based simulations, \texttt{generate\_observation\_scenario\_from\_cpgs()} reads the visit list, updates the roll angle for each visit, selects the target or reference star as appropriate, and appends all generated frames into a single output list. If full-frame output and FITS saving are enabled, each frame can be written as an L1-style FITS product during the sequence generation.

\label{sec:inputs}

\section{VALIDATION AND EXAMPLE USE CASES}
\label{sec:validation}

\subsection{Development Coordination}

\texttt{corgisim} is maintained with Git and hosted on GitHub. The \texttt{main} branch is protected, and changes can be merged only after review and approval. Development goals, bug reports, and feature requests are managed through the GitHub issue tracker, where users are encouraged to report problems and suggest improvements. Contribution guidelines are provided in the repository README. Contributors develop changes on feature branches and submit pull requests to the \texttt{main} branch. Each pull request is reviewed by the project maintainers and evaluated using automated unit tests. Reviewers may also run relevant  tests to verify specific functionality, as discussed in the following section. Development activities are further coordinated through the CPP Simulations teleconferences, which are held every two weeks.

\subsection{Testing and Validation}
\label{subsec:tests}

Validation is carried out at several levels. First, an installation test verifies that the Roman preflight \texttt{PROPER} prescription files are available and that a minimal \texttt{Scene}--\texttt{CorgiOptics}--\texttt{CorgiDetector} workflow can be initialized. For reference validation, corgisim outputs are compared with direct \texttt{CGISim} simulations for standard on-axis star and off-axis companion cases. The tests also check that spectra are scaled to the expected V-band magnitude, that satellite spots appear at the expected locations and contrast, and that polarization channels have the expected relative intensities. Other tests verify roll-angle coordinate conversions, spectroscopy slit/prism behavior, jitter and finite-stellar-diameter weights, detector image generation, and L1-style FITS headers. Together, the continuous integration tests help ensure that standard simulations keep working as the code is updated.

\subsection{Representative Simulation Cases}
\label{subsec:examples}

The current version of \texttt{corgisim} supports imaging, point-source polarimetry, spectroscopy, visit-sequence generation, and full-frame EMCCD simulations, as well as adding pointing jitter and simulating the finite diamater of the host star. Figure~\ref{fig:fig2} presents representative \texttt{corgisim} products. Panels (a) and (b) present two imaging configurations. The Hybrid Lyot Coronagraph (HLC) narrow-field-of-view (NFOV) mode in Band~1 (575\,nm) produces a compact dark-hole region optimized for companions at small angular separations, whereas the Shaped Pupil Coronagraph (SPC) wide-field-of-view (WFOV) mode provides larger spatial coverage for planets at larger separations. Panel (c) demonstrates the wider field and longer wavelength of the SPC Band~4 (825\,nm) mode with a simulation of an extended circumstellar disk. Panel (d) shows HLC polarimetric imaging in Band~1, in which orthogonal 0$^\circ$ and 90$^\circ$ polarization channels enable polarized circumstellar signals to be distinguished from residual stellar light. Panel (e) shows SPC bowtie spectroscopy in Band~3 (730,nm), in which light from the companion passes through a slit and is dispersed by a prism to produce a  spectrum. Panels (f) and (g) show the final \texttt{corgisim} product: a full-frame EMCCD image incorporating detector effects, including dark current, flat-field variations, cosmic-ray events, and read noise. The output is saved as a standardized FITS file ready for processing by the data-reduction pipeline.

\section{FUTURE DEVELOPMENT}

Future development will focus on two areas: time-dependent speckle simulations
for time-series observations and polarimetric propagation of two-dimensional
scenes. The current implementation primarily treats the coronagraphic speckle
field as static, within an observing
sequence. Future versions will add controlled speckle variability in different exposures to support
tests of observing strategies, calibration methods, and time-domain analysis.
However, these simulations are not intended to generate fully accurate
on-the-fly predictions of the true on-orbit speckle evolution, which will depend
on detailed thermal, mechanical, and wavefront-control behavior. Instead, the
goal is to provide realistic and configurable speckle variability for pipeline
and science-performance studies. Additional development will also extend the
polarimetric treatment of two-dimensional scenes, such as disks, so that spatially
extended polarized emission can be propagated more consistently through the
 observing modes.

\section{SUMMARY}
\label{sec:conclusions}

\texttt{corgisim} is a Python simulation framework for the Roman Coronagraph Instrument. It connects astrophysical scene generation, optical propagation with the Roman preflight \texttt{PROPER} model, and EMCCD detector modeling through the \texttt{Scene}, \texttt{CorgiOptics}, and \texttt{CorgiDetector} components. The package supports imaging, polarimetry, and spectroscopy of host stars, companions, and extended sources, together with effects such as telescope roll, finite stellar diameter, pointing jitter, satellite spots, and detector noise.

Simulations may be configured directly in Python or generated from CPGS observing specifications. \texttt{corgisim} produces both noiseless optical subframes and full frame, L1 FITS products with the compatible headers and Roman's official naming conventions. Comparison with \texttt{CGISim} and regression tests help maintain the accuracy and stability of the package. By providing an open-source, end-to-end framework, \texttt{corgisim} supports pre-launch calibration, pipeline development, observation planning, and broader community studies of the Coronagraph Instrument data.

\clearpage
\appendix    
\section{Tables for \texttt{corgisim} input parameters}

\begin{table}[h]
\centering
\small
\caption{Input parameters used by the \texttt{Scene} class to define astrophysical scenes in \texttt{CorgiSim}.}
\label{tab:scene_inputs}
\begin{tabular}{lp{1cm}p{7cm}ll}
\hline
\textbf{Input Key} & \textbf{Type} & \textbf{Description} & \textbf{Status} & \textbf{Default} \\
\hline
\multicolumn{5}{l}{\textbf{\texttt{host\_star\_properties}}} \\
\hline
\texttt{Vmag} & float &
Host star apparent magnitude in the V band. &
Required & -- \\

\texttt{magtype} & str &
Magnitude system, e.g., \texttt{vegamag} or \texttt{ABmag}. &
Required & -- \\

\texttt{spectral\_type} & str &
Host star spectral type, e.g., \texttt{G2V}. &
Required & -- \\

\texttt{target\_name} & str &
Target name used in header metadata. &
Optional & \texttt{UNKNOWN} \\

\texttt{RA} & float &
Target right ascension used in header metadata. &
Optional & \texttt{0.0} \\

\texttt{DEC} & float &
Target declination used in header metadata. &
Optional & \texttt{0.0} \\

\texttt{ref\_flag} & bool &
Reference star flag. &
Optional & \texttt{False} \\

\texttt{stellar\_diam\_mas} & float &
Angular diameter of the host star in mas used for finite stellar diameter simulations. If not provided, the host star is treated as a point source. &
Optional & \texttt{None} \\

\hline
\multicolumn{5}{l}{\textbf{\texttt{companion\_info}}} \\
\hline
\texttt{Vmag} & float &
Companion apparent magnitude in the V band. &
Required & -- \\

\texttt{magtype} & str &
Magnitude system, e.g., \texttt{vegamag} or \texttt{ABmag}.&
Required & -- \\

\texttt{position\_x} & float &
Companion offset relative to the host star in the RA direction, in mas. &
Required & -- \\

\texttt{position\_y} & float &
Companion offset relative to the host star in the DEC direction, in mas. &
Required & -- \\

\texttt{Custom\_Spectrum} & spectrum object &
Custom companion spectrum used instead of the default flat spectrum. &
Optional & \texttt{None} \\

\texttt{Rescale\_Custom\_Spectrum} & bool &
Flag indicating whether the custom spectrum should be rescaled to match \texttt{Vmag}. &
Optional & \texttt{False} \\

\texttt{pol\_state} & array-like &
Companion Stokes vector \texttt{[I,Q,U,V]}. &
Optional & \texttt{[1,0,0,0]} \\
\hline

\multicolumn{5}{l}{\textbf{\texttt{twoD\_scene\_info}}} \\
\hline
\texttt{disk\_model\_path} & str &
Path to the FITS file containing the two-dimensional disk model. The image pixel values are converted to detector count rates according to \texttt{flux\_unit}, or scaled according to \texttt{contrast} when \texttt{flux\_unit} is not provided. &
Required & -- \\

\texttt{prf\_path} & str &
Path to the precomputed, field-dependent point response function (PRF) cube used for field-dependent PRF convolution. &
Required & -- \\

\texttt{contrast} & float &
Average disk surface-brightness contrast relative to the host star, in magnitudes. Used to scale the disk model when \texttt{flux\_unit} is not provided. &
Optional & \texttt{16} \\

\texttt{flux\_unit} & str &
Surface-brightness unit assigned to the disk-model pixel values and used for conversion to detector count rate. Supported units are \(\mathrm{mJy\,arcsec^{-2}}\), \(\mathrm{W\,m^{-2}\,pixel^{-1}}\), and (contrast/pixel). &
Optional & \(\mathrm{mJy\,arcsec^{-2}}\) \\

\hline
\end{tabular}
\end{table}

\begin{center}
\small
\begin{longtable}{p{3cm}lp{8cm}p{1.5cm}p{1cm}}
\caption{Input parameters used by the \texttt{CorgiOptics} class to define the  optical configuration in \texttt{CorgiSim}.}
\label{tab:corgioptics_inputs} \\
\hline
\textbf{Input Key} & \textbf{Type} & \textbf{Description} & \textbf{Status} & \textbf{Default} \\
\hline
\endfirsthead

\multicolumn{5}{l}{\tablename\ \thetable{} -- continued from previous page} \\
\hline
\textbf{Input Key} & \textbf{Type} & \textbf{Description} & \textbf{Status} & \textbf{Default} \\
\hline
\endhead

\hline
\multicolumn{5}{r}{Continued on next page} \\
\endfoot

\hline
\endlastfoot

\hline
\multicolumn{5}{l}{\textbf{Primary inputs}} \\
\hline

\texttt{cgi\_mode} & str &
CGI simulation mode, e.g., \texttt{excam}, \texttt{spec} \texttt{excam\_efield}. &
Required &
\texttt{None} \\

\texttt{bandpass} & str &
Roman CGI bandpass or filter name used to define the wavelength range and throughput. &
Required &
\texttt{None} \\

\texttt{diam} & float &
Telescope primary mirror diameter in cm used by the optical model. &
Optional &
\texttt{236.3114} \\

\texttt{roll\_angle} & float &
Telescope roll angle in degrees. This defines the rotation of EXCAM coordinates relative to sky coordinates. &
Optional &
\texttt{0} \\

\texttt{oversampling\_factor} & int &
Detector oversampling factor when generating optical images before binning to detector pixels. &
Optional &
\texttt{7} \\

\texttt{return\_oversample} & bool &
Flag indicating whether to return the oversampled image instead of the detector-sampled image. &
Optional &
\texttt{False} \\

\texttt{visit\_type} & str &
Visit type used to populate the output FITS header keyword \texttt{VISTYPE}. &
Optional &\texttt{CGI\_VST} \texttt{\_TDD\_OBS}\\

\texttt{visit\_ID} & str &
16 character visit identifier used to populate the output FITS header keyword \texttt{VISITID}. &
Optional &
\texttt{02000010} \texttt{01001001} \texttt{001}\\

\texttt{if\_quiet} & bool &
Flag used to suppress printed output during optics initialization. &
Optional & False \\

\texttt{optics\_keywords} & dict &
Dictionary of keywords passed to the Roman preflight \texttt{PROPER} model. &
Required &
\texttt{None} \\

\texttt{satspot\_keywords} & dict &
Dictionary used to add satellite spots through a DM cosine pattern. &
Optional &
\texttt{None} \\

\texttt{stellar\_diam} \texttt{\_and\_jitter\_keywords} & dict &
Dictionary used to include finite stellar diameter and/or pointing jitter effects. &
Optional &
\texttt{None} \\

\hline
\multicolumn{5}{l}{\textbf{\texttt{optics\_keywords}}} \\
\hline 

\texttt{cor\_type} & str &
Coronagraph type, e.g., \texttt{hlc}, \texttt{spc-wide}, \texttt{spc-spec\_band2}, or \texttt{spc-spec\_band3}. &
Required &
\texttt{hlc} \\

\texttt{polaxis} & int &
Polarization axis used by the optical model. Special values are used for averaged or full polarization-aberration cases. &
Required &
\texttt{0} \\

\texttt{output\_dim} & int &
Output image dimension in pixels before detector placement. &
Required &
\texttt{201} \\

\texttt{use\_errors} & int &
Flag indicating whether optical surface phase errors are included. &
Optional &
\texttt{1} \\

\texttt{use\_fpm} & int &
Flag indicating whether the focal-plane mask is included. &
Optional &
\texttt{1} \\

\texttt{use\_lyot\_stop} & int &
Flag indicating whether the Lyot stop is included. &
Optional &
\texttt{1} \\

\texttt{use\_field\_stop} & int &
Flag indicating whether the field stop is included. &
Optional &
\texttt{1} \\

\texttt{use\_pupil\_lens} & int &
Flag indicating whether the pupil-imaging lens is included. &
Optional &
\texttt{0} \\

\texttt{use\_dm1}, \texttt{use\_dm2} & int &
Flags indicating whether DM1 and DM2 are included in the optical propagation. &
Optional &
\texttt{0} \\

\texttt{dm1\_v}, \texttt{dm2\_v} & array-like &
DM voltage maps used by the optical model. Satellite spots are added by modifying \texttt{dm1\_v}. &
Optional &
\texttt{0} \\

\texttt{fsm\_x\_offset\_mas} & float &
Fast steering mirror offset in mas. &
Optional &
\texttt{0} \\

\texttt{fsm\_y\_offset\_mas} & float &
Fast steering mirror offset in mas. &
Optional &
\texttt{0} \\

\texttt{nd} & int &
Neutral-density filter identifier. Valid values are \texttt{0}, \texttt{1}, \texttt{2}, and \texttt{3}. &
Optional &
\texttt{0} \\

\texttt{prism} & str &
Polarimetric or spectroscopic prism selection. Examples include \texttt{POL0}, \texttt{POL45}, \texttt{PRISM2}, and \texttt{PRISM3}. &
Optional &
\texttt{None} \\

\texttt{slit} & str &
Named FSAM slit used in spectroscopy mode. &
Optional &
\texttt{None} \\

\texttt{slit\_dec\_offset\_mas} & float &
Spectroscopic slit offset in sky coordinates, in mas. &
Optional &
\texttt{0.0} \\

\texttt{wav\_step\_um} & float &
Wavelength step size used by the spectroscopy prism dispersion model, in microns. &
Optional &
\texttt{1e-3} \\

\hline
\multicolumn{5}{l}{\textbf{\texttt{satspot\_keywords}}} \\
\hline

\texttt{num\_pairs} & int &
Number of satellite-spot pairs to add to DM1. &
Required if used &
-- \\

\texttt{sep\_lamD} & float &
Satellite-spot separation from the star, in units of $\lambda/D$. &
Required if used &
-- \\

\texttt{angle\_deg} & array-like &
Position angle(s) of the satellite spots, in degrees. &
Required if used &
-- \\

\texttt{contrast} & float &
Target satellite-spot contrast. &
Required if used &
-- \\

\texttt{wavelength\_m} & float &
Reference wavelength used to generate the DM cosine pattern, in meters. &
Required if used &
-- \\

\texttt{sign} & str &
Sign of the DM cosine pattern. &
Optional &
\texttt{positive} \\

\hline
\multicolumn{5}{l}{\textbf{\texttt{stellar\_diam\_and\_jitter\_keywords}}} \\
\hline

\texttt{use\_finite\_} \texttt{stellar\_diam} & bool &
Flag indicating whether finite stellar diameter effects are included. &
Optional &
\texttt{False} \\

\texttt{add\_jitter} & bool &
Flag indicating whether pointing jitter effects are included. &
Optional &
\texttt{False} \\

\texttt{jitter\_sigmax} & float &
RMS jitter in the $x$ direction, in mas. &
Required for jitter &
-- \\

\texttt{jitter\_sigmay} & float &
RMS jitter in the $y$ direction, in mas. &
Required for jitter &
-- \\

\texttt{outer\_radius\_of} \texttt{\_offset\_circle} & float &
Outer radius of the offset grid used to sample the stellar disk and/or jitter distribution, in mas. &
Required if used &
-- \\

\texttt{N\_rings\_of\_offsets} & int &
Number of radial rings used in the offset-source grid. &
Required if used &
-- \\

\texttt{N\_offsets\_per\_ring} & array-like &
Number of offset points assigned to each radial ring. &
Required if used &
-- \\

\texttt{starting\_offset} \texttt{\_ang\_by\_ring} & array-like &
Starting position angle for the offset points in each ring, in degrees. &
Required if used &
-- \\

\texttt{r\_ring0} & float &
Radius of the innermost offset ring, in mas. &
Required if used &
-- \\

\texttt{dr\_rings} & array-like &
Radial spacing or width of each offset ring, in mas. &
Required if used &
-- \\

\texttt{N\_offsetgrid} & int &
Resolution of the grid used to calculate finite-star and jitter weights. &
Optional &
\texttt{260} \\

\texttt{use\_saved\_deltaE} \texttt{\_and\_weights} & bool &
Option controlling whether the offset electric-field library is calculated or reused. &
Optional &
\texttt{False} \\
\end{longtable}
\end{center}

\begin{table}
\centering
\small
\caption{Input parameters used by the \texttt{CorgiDetector} class to configure the EMCCD detector model in \texttt{CorgiSim}.}
\label{tab:corgidetector_inputs}
\begin{tabular}{llp{7cm}lp{1.5cm}}
\hline
\textbf{Input Key} & \textbf{Type} & \textbf{Description} & \textbf{Status} & \textbf{Default} \\
\hline
\multicolumn{5}{l}{\textbf{Primary inputs}} \\
\hline

\texttt{emccd\_keywords} & dict &
Dictionary of EMCCD detector parameters passed to \texttt{CorgiDetector}. &
Optional &
\texttt{None} \\

\texttt{photon\_counting} & bool &
Flag indicating whether photon-counting mode is used. &
Optional &
\texttt{False} \\

\hline
\multicolumn{5}{l}{\textbf{\texttt{emccd\_keywords}}} \\
\hline

\texttt{em\_gain} & float &
Electron multiplication gain. &
Optional &
\texttt{1000} \\

\texttt{full\_well\_image} & float &
Full-well capacity of the image section, in electrons. &
Optional &
\texttt{60000} \\

\texttt{full\_well\_serial} & float &
Full-well capacity of the serial register, in electrons. &
Optional &
\texttt{100000} \\

\texttt{dark\_rate} & float &
Dark current rate, in $\mathrm{e^{-}\,pix^{-1}\,s^{-1}}$. &
Optional &
\texttt{0.00056} \\

\texttt{cic\_noise} & float &
Clock-induced charge noise, in $\mathrm{e^{-}\,pix^{-1}\,frame^{-1}}$. &
Optional &
\texttt{0.01} \\

\texttt{read\_noise} & float &
Read noise, in $\mathrm{e^{-}\,pix^{-1}\,frame^{-1}}$. &
Optional &
\texttt{100} \\

\texttt{bias} & float &
Detector bias level, in digital numbers (DN). &
Optional &
\texttt{0} \\

\texttt{qe} & float &
Quantum efficiency. In \texttt{emccd\_keywords}, QE is set to unity because the throughput is already included in the throughput calculation. &
Fixed internally &
\texttt{1.0} \\

\texttt{cr\_rate} & float &
Cosmic-ray event rate, in $\mathrm{hits\,cm^{-2}\,s^{-1}}$. A value of \texttt{0} disables cosmic rays; \texttt{5} approximates the L2 environment. &
Optional &
\texttt{5} \\

\texttt{pixel\_pitch} & float &
Detector pixel pitch, in meters. &
Optional &
\texttt{13e-6} \\

\texttt{e\_per\_dn} & float &
Post-multiplied electrons per data number. &
Optional &
\texttt{1.0} \\

\texttt{numel\_gain\_register} & int &
Number of elements in the EM gain register. &
Optional &
\texttt{604} \\

\texttt{nbits} & int &
Number of analog-to-digital converter bits. &
Optional &
\texttt{14} \\

\texttt{use\_traps} & bool &
Flag indicating whether CTI trap effects are simulated. &
Optional &
\texttt{False} \\

\texttt{date4traps} & float &
Decimal year of observation, used only when \texttt{use\_traps=True}. &
Optional &
\texttt{2028.0} \\
\hline
\end{tabular}
\end{table}

\clearpage
\acknowledgments 

This material is based upon work supported by NASA under awards 80NSSC24K0097, 80NSSC24K0216 and
80NSSC24K0087. A. Lau, S. Noiret and E. Choquet acknowledge the support by the European Union (ERC, ESCAPE, project No. 101044152). Views and opinions expressed are, however, those of the author(s) only and do not necessarily reflect those of the European Union or the European Research Council Executive Agency. 
This research was carried out in part at the Jet Propulsion Laboratory, California Institute of Technology, under a contract with the National Aeronautics and Space Administration (80NM0018D0004).

\bibliography{report} 
\bibliographystyle{spiebib} 

\end{document}